\documentstyle[aps,twocolumn]{revtex}
\begin{document}
\title{Photon-graviton conversion in a primordial magnetic field\\ 
and the cosmic microwave background} 
\author{Anal\'\i a N. Cillis and Diego D. Harari} 

\address{Departamento de F{\'\i}sica, Facultad de Ciencias 
Exactas y Naturales\\ Universidad de Buenos Aires \\ Ciudad Universitaria - 
Pab. 1, 1428 Buenos Aires, Argentina} 
\maketitle 

\begin{abstract} 
\tighten 
We reconsider the effects of photon-graviton conversion in a primordial 
magnetic field upon the cosmic microwave 
background radiation. We argue that plasma 
effects make the photon-graviton conversion process
negligible.
\end{abstract} 
\section{Introduction}

A photon that covers a distance $L$ within a transverse, 
homogeneous magnetic field of strength $B$
has a probability of converting into a graviton given by\cite{Gers,Zeldovich} 
\begin{equation}
P\simeq 4\pi GB^2L^2\simeq 8\times 10^{-50}\left ({B\over {\rm Gauss}}
\right )^2\left ({L\over {\rm cm}}\right )^2,
\label{P}
\end{equation}
where $G$ is Newton's constant.

It has recently been suggested\cite{Chen,Magueijo}
that a primordial magnetic field may imprint observable consequences upon the 
cosmic microwave background radiation through photon-graviton conversion.
According to eq. (\ref{P}), a primordial magnetic field
of present value around $10^{-8}$ Gauss, if it already existed
at the time of
decoupling of matter and radiation and was homogeneous over a Hubble radius,
would have induced
a degree-scale anisotropy of
the cosmic microwave background of about $10^{-5}$, of the order of the 
observed value \cite{1deg}.  Although current bounds suggest that a 
cosmological magnetic field, if it exists, has present strength smaller than
around $10^{-9}$ Gauss\cite{Breviews}, photon-graviton conversion
could in principle provide an 
independent method to constrain or eventually detect a primordial cosmological 
magnetic field.

In this article we wish to point out that plasma effects due to the Universe 
residual ionization make the photon-graviton oscillation length much shorter 
than the Hubble radius, and the probability of photon-graviton conversion is 
consequently much smaller than in the absence of free electrons. The effects 
of a primordial magnetic field of present value around $10^{-9}$ Gauss or 
smaller are consequently negligible.

\section{Photon-graviton conversion probability}

The interaction between a gravitational and an electromagnetic field
linearized in the small perturbation $h_{\mu\nu}$
around flat space-time is described, in General Relativity,  by
the term in the action
\begin{equation}
S_{int}=%
{\displaystyle {1 \over 2}}
\displaystyle \int 
h_{\mu \nu }T^{\mu \nu }d^4x  \label{Sint}
\end{equation}
where $T_{\mu \nu }$ is the flat-space energy-momentum tensor 
of the electromagnetic field.

In an external, homogeneous  magnetic field $B$, 
photons and gravitons can convert 
into each other conserving energy and linear momentum. The 
linearized interaction term between electromagnetic and gravitational
plane-waves with the same wave-vector can be written as
\begin{equation}
S_{int}=B\sin \theta 
\displaystyle \int 
\left[ h_{+}E_{\perp }+h_{\times }E_{\parallel }\right] d^4x\ . \label{Sint2}
\end{equation}
Here $E_\parallel$ and  $ E_\perp $ denote respectively 
the component of the electric field 
in the electromagnetic wave that is either parallel or perpendicular
to the plane that contains the direction of propagation and
the external homogeneous magnetic field,
$h_{+}$ and $h_{\times }$ describe two independent polarization
modes of the gravitational wave, in the transverse-traceless gauge,
and $\theta$ is the angle between
the external magnetic field and the  
common direction of propagation of the electromagnetic and gravitational
waves.

From eq. (\ref{Sint2}) the conversion probability 
between photons and gravitons is easily read off. 
Incoming photons with polarization either $\parallel$ or
$\perp$ convert into gravitons 
with the same probability
\begin{equation}
P = 4\pi GB^2L^2\sin^2\theta\ ,
\label{P2}
\end{equation}
the only difference
being the polarization of the resulting graviton.

We wish to point out that precisely because these two independent
states of linear polarization have the same conversion probability,
unpolarized electromagnetic radiation does not become 
linearly polarized
due to photon-graviton conversion
as it propagates through an homogeneous
magnetic field. In this respect, photon-graviton conversion differs
qualitatively from the conversion between photons and pseudoscalar
particles\cite{S,RS}. In the latter case, only $E_\parallel $
mixes with the pseudoscalar field. Photon-pseudoscalar conversion in a 
cosmological magnetic field induces a small degree of linear 
polarization in the cosmic microwave background \cite{HS}.
We conclude, however, and contrary to the claim in ref. \cite{Magueijo},
that photon-graviton conversion does not induce linear polarization in the
cosmic microwave background.

In the presence of a free electron density $n_e$, photons propagate
as if they had an
effective mass equal to the plasma frequency 
$\omega^2_{pl}=4\pi \alpha n_e / m_e$,
where $m_e$ denotes the electron mass and 
$\alpha =\frac{e^2}{4\pi }\sim \frac 1{137}$ is the fine
structure constant. We work  in Heaviside-Lorentz natural units
(in which $1 {\rm Gauss} = 1.95 \times 10^{-2} {\rm eV}^2$).
If the external magnetic field and the electron density are
perfectly homogeneous, there are oscillations between
the electromagnetic and gravitational plane waves, over an
oscillation length given by\cite{RS} 
\begin{equation}
\ell_{\rm osc}=%
{\displaystyle {4\pi \omega  \over \omega _{pl}^2}}
\label{losc}
\end{equation}
where $\omega $ is the angular frequency of the electromagnetic wave. 
Indeed, the photon-graviton conversion probability, for either 
$\parallel$ or $\perp$ polarization, becomes\cite{RS}
\begin{equation}
P={4\over\pi} GB^2%
\ell_{\rm osc}^2
\sin ^2%
{\displaystyle {\pi L \overwithdelims() \ell_{\rm osc} }}
\sin ^2\theta   \label{P3}
\end{equation}
Of course, if $L\ll l_{\rm osc}$  this expression reduces to
eq. (\ref{P2}), as if there were no free electrons.
Otherwise, the conversion probability does not accumulate 
over distances larger then $\ell_{\rm osc}$.

The situation is different when there are processes, such as
inhomogeneities in the electron-density,
that affect the coherence of the photon-graviton oscillations.
In this case a fraction $f$ of the photons that mixed into gravitons
within one oscillation
does not oscillate back into photons. 
Adding the effect over $N=L/l_{\rm osc}$ independent regions
the photon-graviton conversion probability 
over a distance $L$ becomes
\begin{equation}
P\simeq f GB^2L\ell _{\rm osc} \sin ^2 \theta \label{PH}
\end{equation}
The precise value of the factor $f$ is model-dependent. 
See for instance ref. \cite{CG} for an estimate of these effects 
in the interstellar medium
in our galaxy. For our purposes it will be enough to consider
its largest possible value, $f\simeq 1$.
We shall see that even in this most favourable case, photon-graviton
conversion in a primordial magnetic field has negligible effects.

\section{CMB anisotropy induced by photon-graviton conversion}

Photon-graviton conversion in a cosmological magnetic field
induces anisotropies in the CMB due to the angular dependence of
the conversion probability \cite{Zeldovich,Chen,Magueijo}.
Ignoring plasma effects, the conversion probability is frequency-independent, 
and thus preserves the black-body CMB spectrum.
Using eq. (\ref{P2}) we see that a 
cosmological magnetic field of present value $B(t_\circ)$ 
assumed homogeneous over a scale of order the present Hubble
radius, $H_\circ^{-1}$, would induce (if plasma effects were negligible)
a large angular scale anisotropy of order
\begin{equation}
{\Delta T \over T}
\simeq 5\times 10^{-6}%
{\displaystyle {B(t_{\circ }) \overwithdelims() 1.3\times 10^{-6} 
{\rm Gauss}}}^2%
{\displaystyle {h \overwithdelims() 0.5}}
^{-2}
\end{equation}
where $H_{\circ
}=100\ h\ {\rm km}\ {\rm seg}^{-1}{\rm Mpc}^{-1}.$ 
The anisotropy induced at present times by a cosmological magnetic field
of about $10^{-9}$ Gauss
would thus be negligible, about six orders of magnitude smaller
than the observed quadrupole CMB anisotropy \cite{cobe}, even 
in the absence of plasma effects. 

A cosmological magnetic field of present value $B(t_\circ)$ 
is expected to have been larger in the past, by a factor
$B(t)=B(t_\circ) a^2(t_\circ)/a^2(t)$, where $a$ is the Robertson-Walker
scale factor, due to flux conservation \cite{Breviews}.
Photon-graviton conversion would thus have had larger effects in the past,
if the magnetic field was always homogenous over a Hubble radius,
since the factor $(BH^{-1})^2$ scales with redshift as $1+z$
in a matter-dominated universe.
Anisotropies induced before decoupling, however, are quickly erased
by Thomson scattering during the period of tight coupling
between photons, electrons and baryons. The largest effect would thus
arise right around decoupling.
The anisotropy induced around 
the time of decoupling of matter and radiation ($t=t_*$),
on angular scales of order the size of the horizon at
decoupling, which corresponds to about one degree on our sky is,
neglecting plasma effects
\begin{equation}
 {\Delta T \over T} \approx 10^{-5}%
{\displaystyle {B(t_{*}) \overwithdelims() 0.04 {\rm Gauss}}}^2%
{\displaystyle {h \overwithdelims() 0.5}}
^{-2}%
{\displaystyle {1+z_{*} \overwithdelims() 1100}}
^{-3}\ .
\end{equation}
$10^{-5}$ 
is the order of the observed anisotropy on angular scales of
about one degree\cite{1deg}. The present value of
a primordial magnetic field which had a strength $B(t_*)\simeq 0.04$ Gauss 
at decoupling is
$B(t_{\circ })\simeq 3\times 10^{-8}$ Gauss.
We thus conclude, as in refs. \cite{Chen,Magueijo}, 
that if plasma effects were negligible
the conversion between photons and gravitons in a primordial magnetic field
around the time of decoupling of matter and radiation
could have non-negligible effects upon the isotropy of the CMB.

Plasma effects, however, are not negligible. Even in the most favourable
case, with $f\approx 1$ in eq. (\ref{PH}),
the conversion probability drops precipitously. 
Consider the Universe right after decoupling of
matter and radiation. The number density of free electrons is
\begin{equation}
n_e(t\approx t_*)=0.15%
{\displaystyle {\Omega _bh^2 \overwithdelims() 0.01}}
{\displaystyle {1+z_{*} \overwithdelims() 1100}}^3%
{\displaystyle {X \overwithdelims() 10^{-3}}} {\rm cm}^{-3}
\end{equation}
where $X$ is the fractional residual ionization and $\Omega_b$ is the baryon 
energy-density in units of the critical density. 
Notice that because the oscillation length depends on the photon
frequency, so does the conversion probability. Photon-graviton conversion does 
not preserve the black body spectrum of the CMB. We still write, for 
comparison purposes, the anisotropy in the CMB intensity induced by
photon-graviton conversion in terms of an effective temperature
anisotropy, at a given frequency. The anisotropy induced by a magnetic
field homogeneous over a Hubble radius at decoupling would be at most, 
including plasma effects
\begin{eqnarray}
 {\Delta T \over T}
 &\lesssim &GB^2(t_{*})H^{-1}(t_{*})\ell _{\rm osc}(t_*)  \nonumber
\\
\  &\simeq &10^{-5}%
{\displaystyle {B(t_{*}) \overwithdelims() 14 {\rm Gauss}}}^2%
{\displaystyle {h \overwithdelims() 0.5}}^{-1}%
{\displaystyle {\nu (t_{\circ }) \overwithdelims() 90 {\rm GHz}}}\ 
\nonumber \\
\end{eqnarray}
Here $\nu (t_\circ)$ is the present value of the CMB photons'
frequency. A magnetic field  of strength 14 Gauss at decoupling
would have a strength of order $10^{-5}$ Gauss today.
A realistic value, smaller than $10^{-9}$ Gauss today, would
thus induce anisotropies through photon-graviton conversion 
at least eight orders of magnitude smaller than those observed.

We have already seen that the large angular scale anisotropy in the CMB 
induced at present times by a field of $10^{-9}$ Gauss would be negligible 
even in the absence of free electrons. A small electron-density would
reduce the effect of photon-graviton conversion even further.
The present value of the free electron density in the intergalactic
medium is not known with certainty. The Gunn-Peterson limit on
the abundance of neutral Hydrogen \cite{Steidel} 
however, 
suggests that most of the intergalactic material is ionized. 
A probably realistic figure for the present electron number density
is thus $n_e\approx 10^{-7} {\rm cm}^{-3}$. 
The anisotropy induced today on large angular scales 
by a cosmological magnetic field would thus be
\begin{eqnarray}
{\Delta T \over T}
 &\lesssim &5\times 10^{-6}%
{\displaystyle {B(t_{\circ }) \overwithdelims() 0.006 {\rm  Gauss}}}
^2  \nonumber \\
&&\ \ \times 
{\displaystyle {\nu  \overwithdelims() 90 {\rm GHz}}}
{\displaystyle {10^{-7}{\rm cm}^{-3} \overwithdelims() n_e}}
{\displaystyle {h \overwithdelims() 0.5}}
^{-1}
\end{eqnarray}
Clearly, the effect of a cosmological magnetic field of
current value around $10^{-9}$ Gauss would be completely negligible.

\section{Conclusions\label{conclus}}

Photon-graviton conversion induced by a cosmological magnetic field
of present strength $10^{-9}$ Gauss or smaller has negligible
effects upon the isotropy of the cosmic microwave background.
The effect would have been much larger in the absence of free electrons.
Plasma effects, however, make the characteristic length for photon-graviton
oscillations much smaller than the Hubble radius, preventing the conversion
probability to grow quadratically with distance over such large scales.

We have also seen that photon-graviton conversion does not induce linear 
polarization upon the cosmic microwave background,
contrary to the case of photon-pseudoscalar
conversion \cite{HS,CG}.

The probability of photon-graviton conversion in a magnetic field 
in the presence of free electrons depends on the photon frequency.
In principle, one could also attempt to detect the effects of a primordial
magnetic field through departures from the black body spectrum in the CMB.
Since thermalization processes are effective only at redshifts larger
than about $z\approx 10^6$, one could test in this way for the presence of
a primordial magnetic fields at very early times. One can show, however,
that the departure from the black body spectrum is also negligibly
small. For a present field of $10^{-9}$ Gauss at 
CMB frequencies of order $10^3$ GHz,
the fractional departure from a black body spectrum is
at most of order $10^{-12}$, induced right after decoupling.
At earlier times, with matter fully ionized,
the large free electron density makes the effect
even smaller, of order $10^{-16}$ at the time of matter-radiation
equality,  $z_{\rm eq}\approx 10^4$. At higher redshifts
the factor $B^2H^{-1}\ell_{\rm osc}$ remains constant.

We should mention that a primordial magnetic field may still have
significant effects upon the isotropy of the cosmic microwave background
by driving an anisotropic expansion of the Universe \cite{Zeldovich}. 
Its direct effect through 
photon-graviton conversion, however, is negligible due to plasma effects.

\section*{Acknowledgements}

This work was partially supported by grants from Universidad de Buenos
Aires and Fundaci\'on Antorchas. D.H. is also supported by CONICET.

\end{document}